# On the mathematics of beauty: beautiful music

A. M. Khalili[1]

*Abstract*—**The question of beauty has inspired philosophers and scientists for centuries, the study of aesthetics today is an active research area in many fields. In this paper, we will study the simplest kind of beauty that can be found in a simple piece of music and can be appreciated universally. The proposed model suggests that beautiful music is the result of an optimization process between randomness and regularity. Then we show that beautiful music delivers higher amount of information over multiple levels in comparison with less aesthetically appealing patterns when the same amount of energy is used. The proposed model is tested on a set of beautiful music pieces.**

***Index Terms*—Evolutionary Art, Music Aesthetic Assessment, Information Theory.**

## I. Introduction

The study of aesthetics started with the works of ancient Greek and recently became an active research area in many fields. Predicting the aesthetic appeal of music is beneficial for a number of applications, such as retrieval and recommendation in multimedia systems. Recently, convolutional neural networks (CNN), which can automatically learn the aesthetic features have been applied to aesthetic quality assessment [8], [9], [10], and [11], promising results were reported. Neural networks have also been used in music composition, some early approaches include [12], [13], [14], [15], [16], [17], [18], and [19]. The development of a model of aesthetic judgement is also one of the major challenges in evolutionary art [20], [21], and [22] where only music of high aesthetic quality should be generated.

Birkhoff [23] proposed an information theory approach to aesthetics, he proposed a mathematical measure, where the measure of aesthetic quality M = O/C is in direct relation to the degree of order O, and in reverse relation to the complexity C. Eysenck [24], [25], and [26] suggested that the aesthetic measure function should be in a direct relation to complexity rather than an inverse relation M = O×C. Javid et al. [27] conducted a survey on the application of entropy to quantify order and complexity, they proposed a computational measure of complexity, the model is based on the information gain from specifying the spacial distribution of pixels and their uniformity and non-uniformity. Herbert Franke [28] argued that artists should produce a flow of information of about 16 bits per second for their works of art to be perceived as beautiful and harmonious.

Manaris et al. [29], Investigated Arnheim's view [30], [31], and [32] that artists tend to produce art which create a balance between chaos and monotony. They presented results of applying Zipf's Law to music. They created a large set of metrics based on Zipf's Law which measure the distribution of various parameters in music, such as pitch, duration, harmonic consonance, and melodic intervals. They applied these metrics to a large collection of music pieces, their results suggest that metrics based on Zipf's Law, capture essential aspect of proportion in music as it relates to music aesthetic. Simple Zipf metrics have a key limitation, they examine the music piece as a whole, and ignore potentially significant details. For example, sorting a piece's notes in different order of pitch would produce an unpleasant musical artefact, this artefact exhibits the same distribution as the original piece. Therefore Fractal metrics were used in [29] to handle the limitation of simple metric, the fractal metric will capture how many subdivisions of the piece exhibit the same distribution at many levels of granularity. For example, they recursively applied the simple pitch metric to the piece's half subdivisions, quarter subdivisions, etc. However, as stated by the authors, this condition is a necessary but not sufficient condition.

In this paper we will propose a novel model to classify and generate beautiful music. The main contribution of this paper is showing that aesthetically appealing patterns deliver higher amount of information over multiple levels in comparison with less aesthetically appealing patterns when the same amount of energy is used. The paper is organised as follows: Section II describes the proposed model. The results are listed in Section III, and the paper is concluded in Section IV.

## II. Proposed model

In this paper, we will study the simplest kind of beauty that can be found in a short piece of music and can be appreciated universally. Our brain is able to classify very short piece of music and determine whether it is beautiful or not. Therefore, we will not study the whole pieces of music; we will only study the most aesthetically pleasing part of the piece. The most important aspect in identifying beautiful music is by studying the spacing or the transition pattern between different notes, i.e. the difference between the frequencies of successive notes. For simplicity, the frequencies of the studied pieces will be scaled down to a scale close to the human voice. By plotting the transition patterns of different beautiful pieces, a distribution similar to Maxwell-Boltzmann distribution shows up. Fig. 1 shows the distribution of the transitions of parts of three beautiful pieces. However the distribution alone cannot capture the spatial arrangement of the piece; therefore, we will use the multilevel approach to represent the spatial arrangement of the piece.

[1]A. M. Khalili is with the Faculty of Computing, Engineering and Science, Staffordshire University, United Kingdom, (e-mail: a.m.khalili@outlook.com).

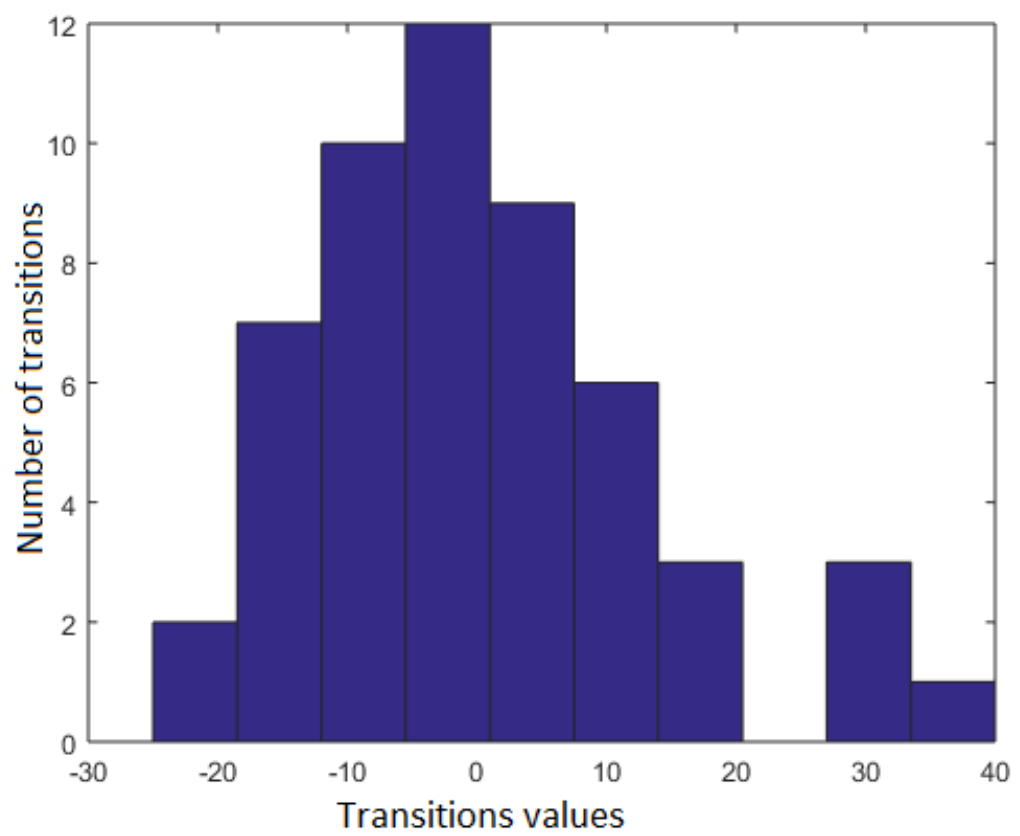

Fig. 1. The distribution of the transition pattern of three beautiful pieces.

If we take the following piece $L_1$ = [120, 160, 170, 145], the values here represent the frequencies of the notes, the corresponding transition pattern will be $L_2$ = [40, 10, -25] which can be obtained by subtracting the notes. Finally, the third level representation for the previous piece will be $L_3$ = [30, 35] which can be obtained by subtracting the value of the second level.

We will use the same model proposed in [34], the entropy will be used as a measure of randomness, and energy as a measure of regularity. Statistical metric such as entropy cannot distinguish between different arrangements of the piece, because it is based on the distribution of the values, and not on their spatial arrangement. Therefore, we will use the multilevel approach to represent the spatial arrangement of the piece, where the first level will represent the frequencies of the notes, the second level represents the transition pattern of the notes, and the third level will represent the difference of the transition pattern of the notes.

In [34] we showed that aesthetically appealing patterns deliver higher amount of information over multiple levels in comparison with less aesthetically appealing patterns when the same amount of energy is used. Where beautiful patterns have a balance between randomness and regularity, and aesthetically appealing patterns are those which are closer to this optimal point. The measure of aesthetic quality M is given by (1)

$$\mathrm{M} = \sum_{\mathrm{i}} \mathrm{Entropy(Li)} \qquad (1)$$

$L_1$, $L_2$, and $L_3$ represent the levels described earlier. Entropy is Shannon entropy. Energy is the sum of the values of $L_i$ and it is given by (2)

$$Energy(Li) = \sum_{n} Li(n) \qquad (2)$$

Maximising the entropy when the energy is the same can then be seen as delivering the highest amount of information using the same amount of energy. Similar to [34], pieces will only be compared with other pieces in the same energy level, i.e. the energy of the three level should be the same.

## III. Results

To test the proposed model, five pieces with four notes will be considered. The pieces with their M value are shown in Table 1. Fig. 2 shows a visual representation of the pieces, where the x-axis represents the duration of the notes, and the y access represents the frequency of the notes.

After that, we generated all possible combinations of the each piece, i.e. we generated all possible pieces that have the same energy at the three levels of the piece, and kept only those that have larger or equal M value.

$P_1$, $P_4$, and $P_5$ showed the highest M value at that energy. For $P_2$ three pieces showed higher M value, and for $P_3$ three pieces showed higher M value; however the pieces found to be aesthetically appealing.

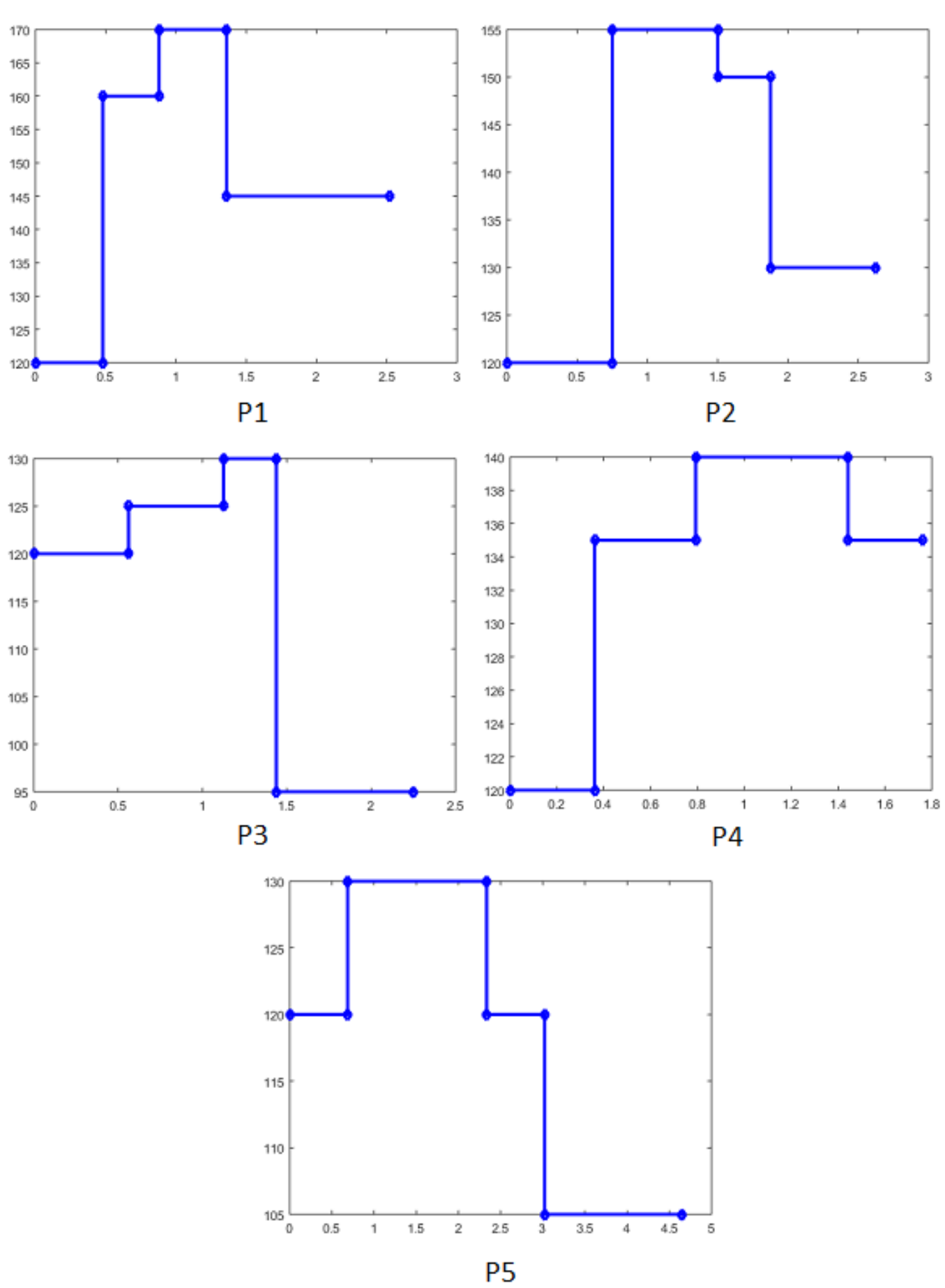

Fig. 2. Visual representation of the five pieces.

Table. 1. The tested pieces with their M value.

| Piece | M value |
|---|---|
| P1 = [120, 160, 170, 145] | 3.584 |
| P2 = [120, 155, 150, 130] | 2.918 |
| P3 = [120, 125, 130, 95] | 2.918 |
| P4 = [120, 135, 140, 135] | 3.084 |
| P5 = [120, 125, 120, 105] | 3.084 |

## IV. Conclusion

This paper has presented a novel approach in identifying beautiful music. The proposed model showed that there is a link between beautiful music and an optimisation process that seeks to deliver the highest amount of information using the same amount of energy. The results showed that the proposed model was able to classify beautiful music.